\documentclass[twocolumn,showpacs,nofootinbib]{revtex4}

\usepackage{epsf}
\usepackage{graphicx}
\usepackage{dcolumn}

\def\be{\begin{equation}}
\def\ee{\end{equation}}
\def\bea{\begin{eqnarray}}
\def\eea{\end{eqnarray}}

\begin{document}
\input epsf
\draft
\renewcommand{\topfraction}{0.8}
\preprint{hep-th/0304225, \today}
\title {\Large\bf M/String Theory, S-branes and Accelerating Universe }
\author{ Michael Gutperle} 
\affiliation{Department of Physics and Astronomy, UCLA, Los Angeles,
  CA}
\affiliation{  Department
  of Physics, Stanford University, Stanford, CA 94305-4060,
USA}  
  
\author{ Renata Kallosh}
\affiliation{  Department
  of Physics, Stanford University, Stanford, CA 94305-4060,
USA}  
  
\author{Andrei Linde}
\affiliation{  Department
  of Physics, Stanford University, Stanford, CA 94305-4060,
USA}  
  
{\begin{abstract}
Recently it was observed that the hyperbolic compactification of
M/string theory related to S-branes may lead to a transient period of
acceleration of the universe.   
We study  time evolution of the corresponding effective 4d
cosmological model supplemented by cold dark matter 
and show that it is marginally possible to describe observational data
for the late-time cosmic acceleration in this model.  
However, investigation  of the compactification $11d \to 4d$ suggests
that the Compton wavelengths  of the KK modes in this model are of the
same order as the size of the observable part of the universe.  
Assuming that this problem, as well as several other problems of this
scenario, can be resolved,  
we propose a possible solution of the cosmological coincidence problem
due to relation between the dark energy density and the effective
dimensionality of the universe. 
\end{abstract}}
\maketitle 

\section{Introduction}

The problem of understanding the origin of a dark energy in the
framework of the fundamental M/string theory after the recent WMAP
data \cite{Bennett:2003bz} became even more urgent than before:  the
basic conclusion from all previous observations that  $\sim 70 \%$ of
the energy density of the universe is in a dark energy sector has been
confirmed. This implies the existence of a stage of late-time
acceleration of the universe. Moreover, the possibility of the early
universe acceleration (inflation) 
is also supported by the data. Thus one would like to derive the
accelerated 4-dimensional universe from the fundamental
11/10-dimensional M/string theories. 

Recently  models of the compactified non-perturbative string theory
have been found which have  de Sitter space vacua
\cite{Kachru:2003aw}. These models give us a possibility to describe  
the  late-time acceleration in the non-perturbative string theory:
the corresponding  equation of state $w=-1$  reflects the fact that a
de Sitter minimum is attained and we have a small positive
cosmological constant. 

Another interesting new development is based on the study of the
time-dependent M/string theory solutions, S-branes
\cite{Gutperle:2002ai}-\cite{sbunch} with hyperbolic 
compactification of the internal 7/6-dimensional space
\cite{TW}-\cite{Chen:2003ij}. It has been discovered in \cite{TW}
that some solutions of 11d supergravity describe a period of
accelerated 4-dimensional 
cosmology. It was pointed out in \cite{ohta} that S-brane solutions in
a more general case may also lead to accelerating cosmologies.  A
natural question arises: what is the equation of state 
$w=p/\rho$ for the dark energy in these models. Here it is important
to stress that this function $w$ may depend on time $t$, or redshift $z$. The
most recent observational constraint on $w$ from WMAP+ supernova
requires that  $w< -0.78$. However, this observational constraint is
valid only under assumption that $w=\rm const$. 

An interpretation of the S-brane solutions with accelerating
cosmologies was given in \cite{Emparan:2003gg}: a 4-dimensional
effective action was presented there whose time dependent solutions
are exactly the time dependent solutions of the 11-dimensional
supergravity with  time-dependent compactification. The corresponding
4-dimensional cosmology was studied in \cite{Emparan:2003gg}
in the setting where the total cosmological evolution at all time is
affected only by the dark energy stress-tensor, $\Omega_D =1$. We
would like to make such models more realistic, i. e. to add matter to 
describe the observed universe   which was matter dominated in the
past, and which presently  has  $\Omega_D\equiv {\rho_D\over \rho_{\rm
    total}}\approx 0.7$ and $\Omega_M \equiv {\rho_M\over \rho_{\rm
    total}}\approx 
0.3$.  
For this purpose one should find out where the matter comes from in
the compact hyperbolic compactifications.  One may think of several
possibilities: 

1. There is a brane world construction, where one adds a space filling 
3-brane which carries the matter fields, as suggested in \cite{hyperb}. 

2. The compact hyperbolic space $H_n/\Gamma$ is basically given by a freely 
acting orbifold (no fixed points). One could also consider orbifolds which 
have fixed points. In the discussion of M-theory on $G_2$ manifolds,
matter is localized at singularities of the $G_2$ manifold
\cite{Atiyah:2001qf}-\cite{Acharya:2001gy}. The problem is that in the
hyperbolic context  
in M-theory one does not have any clear  idea what happens at the orbifold
singularities.

3. One could consider CY spaces with metrics which have negative 
curvature. 

At present none of these possibilities is  clear and  moreover there
are conceptual problems in introducing 4d cold dark matter starting
with  higher dimensional theory \cite{Emparan:2003gg}.  
Here we propose to reinterpret the S-brane models of
\cite{TW}-\cite{Chen:2003ij} as follows: from the M/string theory
we find out the relevant exponential scalar field potentials for the effective
4-dimensional cosmologies. Such potentials serve as a part of the
effective cosmological model describing the late-time accelerating
universe. Then we introduce density of matter phenomenologically, by
adding to the Friedmann equations the energy density $\rho_M$
decreasing as $a^{-3}(t)$, where $a(t)$ is the scale factor. The
resulting Friedmann equations are different from the pure dark energy 
evolution (or S-brane solutions of 11-dimensional theory). We will
find the solutions of these equations numerically and identify
``today'' with the point in time 
when $\Omega_D\sim 0.7$.   This will also allow us to calculate the
relevant evolution of the equation of state for the M-theory type dark
energy, see Section III. 

In fact, the calculations of precisely this type have been already
performed in
\cite{Kallosh:2002gf}, see Section V on ``M-theory and Dark Energy
with Exponential Potentials''. 
Now the new S-brane type models give a specific combination of the
exponential potentials. We will study these 
models following the methods developed in \cite{Kallosh:2002gf} and we
will find equation of state $w(t) = p_D/\rho_D$ for the new models. 

This approach can be quite reasonable from the point of view of the
effective 4d theory. However, as we will see, investigation  of the
compactification $11d \to 4d$ suggests that the Compton wavelengths
$m_{\rm KK}^{-1}$ of the Kaluza-Klein modes in this model are of the
same order as the size of the observable part of the universe. If this
is the case, our universe would be effectively 11d, see Section II.  
On the other hand, if one finds a way to make the KK modes heavy, one
may encounter large quantum corrections $O(m_{\rm KK}^4)$ to the
cosmological constant. These problems may completely rule out this
model.  

Nevertheless, the models of this type have some interesting features
which may deserve further investigation. In particular, in Section IV
we will show that 
if the problems mentioned above can be solved, one can simultaneously
find a solution of the cosmological coincidence problem due to
relation between the dark energy density and the effective
dimensionality of the universe.

\section{The model}

For M-theory case one starts with  11-dimensional Lagrangian 
\be\label{M}
I=\frac{1}{16\pi G_{11}}\int d^{11}x \sqrt{-G} \left( R[G]
-\frac{1}{2\times 4!} F_{[4]}^2
\right) .
\ee
The time dependent solution is  a warped product of a four-dimensional
spacetime 
and an internal compact space  with $R_{ab}(\Sigma)=-
6 g_{ab}r^{-2}_c$, where $r_c$ is the radius of curvature of the
internal space:
\be\label{compac}
ds^2=e^{-7M(x)} g_{\mu\nu}(x)dx^\mu
dx^\nu+e^{2M(x)}d\Sigma^2\,.
\ee
The field strength is taken as $\ast F_{[4]}= b \,{\rm
vol}(\Sigma)$ and $g_{\mu\nu}$ is the Einstein metric in four
dimensions. Upon the dimensional reduction we get \cite{Emparan:2003gg}
\be\label{effective}
I=\frac{1}{16\pi G_{4}}\int d^4x \sqrt{-g}\left( R[g]
-\frac{63}{2}(\partial M)^2 -
2V(M)\right)\, ,
\ee
where \cite{E1}
\be\label{psipot}
V(M)= \frac{b^2}{4}\; e^{-21M}+
21\,e^{-9M}r^{-2}_c\, .
\ee
The relation between the gravitational Newton constants in 11 and 4
dimensions  is
\be\label{plancka}
 {G_{11}}={V_7\,  G_{4}} \, ,
\ee
where $V_7$ is a constant time independent volume of the internal
compact space $\Sigma$. 
The 4-dimensional metric defining the 11-dimensional solution is given by
\be
g_{\mu\nu}(x)dx^\mu
dx^\nu= -S^6(\xi) d\xi^2+ S^2(\xi) d\vec x^2 \ .
\ee
Using $t$ as a proper time  of the 4d observer with $ dt= \pm S^3(\xi)
d\xi$ it becomes 
\be
g_{\mu\nu}(x)dx^\mu
dx^\nu= - dt^2+ S^2(\xi(t)) d\vec x^2  \ .
\ee
Note that we are using a notation $t$ for the proper time of the 4d
observer to avoid a confusion which notation of  \cite{TW} may cause,
where the proper time of the 4d observer is called $\eta$. In standard
FRW cosmology $\eta$ is reserved for a conformal time $ds^2=
a^2(\eta)(-d\eta^2+ d\vec x^2)$. 

We may change variables so that ${63\over 2}M^2=\left({\phi\over
    M_{Pl}}\right)^{2}$ and 

\be\label{effective2}
L=\frac{M^2_{Pl}}{2}\, R[g]
-\frac{1}{2}(\partial \phi)^2 - V[\phi]
\ ,
\ee
where
\be
V[\phi]=\frac{b^2M^2_{Pl}}{4}\; e^{-\sqrt{14}{\phi\over M_{Pl}}}-
21 e^{-\sqrt{18\over 7}{\phi\over M_{Pl}}}\left ({M_{Pl}\over  r_c}\right)^{2}
\ee
We will now fix the Planck mass $M^2_{Pl}=(8\pi G_{4})^{-1}=1$ and
find a canonical form of the 4-dimensional Lagrangian, 
\be\label{effective3}
\frac{1}{2}R[g]
-\frac{1}{2}(\partial \phi)^2 -
\frac{b^2}{4}\; e^{-\sqrt{14}\phi}-
21\, e^{-\sqrt{18\over 7}\phi} r_c^{-2}\ ,
\ee

The solution of  equations following from this 4d Lagrangian is the
same as the one from 11d. 
\be
S(\xi)= \left( {b\, \cosh (3(\xi-\xi_0))\over 3}\right)^{1\over 4}
\left({\sqrt{3\over28}\over  \sinh( 6\sqrt{3\over28}
  |\xi|)}\right)^{{7\over 12}},
\ee
\bea
 \qquad \phi(\xi)&=&\sqrt{63\over 72}\Bigl( \ln\Bigl( {b\, \cosh
   (3(\xi-\xi_0))\over 3}\Bigr) \nonumber\\
&&-\ln \Bigl(\sqrt{28\over 3}{ \sinh( 6\sqrt{3\over28} |\xi| ) }\Bigr)\Bigl)\ .
\eea
The potential for this model has two exponential terms $\sim
e^{\lambda \phi}$ with 
$\lambda^2= 14$ and $\lambda^2= {18\over 7}\approx 2.57$. One expects
to find the 
current value of the dark energy from the potential 
\be\label{dark}
V(\phi) = \frac{b^2}{4}\; e^{-\sqrt{14}\phi}+
21\, e^{-\sqrt{18\over 7}\phi}r_c^{-2}\approx 10^{-120} \ .
\ee
The contribution from the 4-form field with $\lambda^2= 14$ is way too
steep. The only possibility to have a reasonable late-time cosmology
is in a regime where this exponent is small, for example at very
large positive values of $\phi$ assuming that the factor $b$,
specifying the form field, is not huge. In this case, which is
equivalent to the absence of the 4-form field contribution, we may now
study the case $\lambda^2\approx 2.57$ \cite{TW}. Note that only the
hyperbolic compactification is relevant to get the positive potential
in this case. 
The second exponent has a factor $\sqrt{18\over 7}\approx 1.6$. It was
explained in \cite{LopesFranca:2002ek,Kallosh:2002gf} that
for  $\lambda \lesssim 1.7$ a reasonable description of data
is possible with fine-tuned initial conditions.

Thus it is not impossible to use  M-theory motivated
potentials for dark energy. This would mean that the term
$21\,e^{-9M}r^{-2}_c$   in Eq. (\ref{dark}) is of the order
$10^{-120}$ in 4d Plank units.  This raises the 
following concern:  What prevents us to see the extra 7 dimensions?
In particular, we would like  to  establish whether the Kaluza-Klein
reduction of the 
11-dimensional metric (\ref{compac}) and its interpretation as a 4d
cosmology is internally consistent. 

The relation between the curvature radius
$r_c$ and the volume of the seven dimensional manifold $H_7/\Gamma$ is given by
$V_7 =r_c^7 e^\alpha$, where $\alpha$ depends on the
topology of the manifold  \cite{hyperb}. The mass of the massive
Kaluza-Klein modes is determined by the spectrum of Laplace operators
on the seven dimensional manifold. It is believed \cite{hyperb} that
there is a mass gap, and the masses are bounded below by  
\be
m_{\rm KK}  = c\, e^{-9M/2} r_c^{-1} \ ,
\label{KK}\ee
where $c=O(1)$ is some constant. In our model, this  has an important
impact. For $V(M) \sim 21 e^{-9M}r_c^{-2}$, the present value of the
Hubble constant is given by $H^2 \approx V/3 \sim 7 e^{-9M}r_c^{-2}$,
so the mass gap is  
\be
m_{\rm KK}  \sim \sqrt V \sim H  \sim 10^{-60}  \ .
\ee
This means that the mass gap is practically nonexistent, the Compton
wavelengths of KK modes are the same as  the size of the cosmological
horizon, and therefore particles should be able to freely move in all
10 dimensions of space.  

The only possible loophole in this argument that we are able to see
is to assume that one can find spaces with $V_7 =r_c^7 e^\alpha \ll
r_c^7$, i.e. with $e^\alpha\ll 1$. In this case it might happen that
the KK masses will be given by $V_7^{-1/7} =r_c^{-1}
e^{-\alpha/7}$. To make this mass scale greater than 1 TeV one would
need to have an incredibly small volume  $V_7 \lesssim 10^{-315}\,
r_c^7$. In fact, what really matters  is the bound  on the lowest
eigenvalue of the Laplace operators on the seven dimensional
compactified hyperbolic  manifold.   Not much is known about it at
present,  
and the  estimate $V_7^{-1/7}$ for the KK mass in such spaces is very
speculative. Anyway, it seems unlikely that the lowest level of the KK
mass in such models is phenomenologically viable \cite{E2}.

Even if we were able to find a way to ensure the large values for the
KK masses, we would need to encounter another problem related to these
masses. 
There can be an additional contribution to the potential for the
scalar field $\phi$ in (\ref{dark}) coming from the Casimir
energy of the compact space $H_n/\Gamma$
\cite{Appelquist:1982zs,Appelquist:1983vs}. For a manifold
which breaks supersymmetry the contribution is generically of order
$\lambda_s^4$, where $\lambda_s$ is the supersymmetry breaking
scale. In our case this term would be proportional to  $m_{\rm
  KK}^4$. For $m_{\rm KK} \gtrsim 1$ TeV, this term would be 60 orders
of magnitude greater than the present value of dark energy $\sim
10^{-120}$.  

In \cite{Kehagias:2000dg,Nasri:2002rx} it was noted that in principle
odd dimensional compact hyperbolic orbifolds could exist which
preserve some Killing spinors. It would be very interesting to 
check this claim by constructing a concrete example. Furthermore there
are no calculations of the Casimir energy, going beyond
the dimensional analysis given above, for hyperbolic
orbifolds in  more than three dimensions.  For the rest of the paper
we will assume that the Casimir energy is vanishing due to existence
of Killing spinors 
or small compared 
to the dominant term in the potential  (\ref{dark}). However, one
should remember that if this term is not negligible, it can completely
invalidate the model. 
 
Thus we tend to conclude that the model is not really working for the
 purpose of describing current acceleration of the universe in a way,
 consistent with the fundamental eleven-dimensional M-theory. It has
 an  advantage in principle over the models of eleven-dimensional
 supergravity with ``non-compactification'' \cite{Hull:jw} 
 whose 4d cosmology was studied in \cite{Kallosh:2002gf}. This
 advantage, the existence of the finite gap for KK states, is
 invalidated by the extraordinary small value of the dark energy,
 which the model was designed to explain. Still, as in case of N=8
 gauged supergravity  studied in \cite{Kallosh:2002gf} we may study
 the 4d cosmology of the 4d model in eq. (\ref{effective3}) since it
 is derived from  M-theory and may therefore inherit some of its
 properties.

\section{Accelerating Universe}

\subsection{Dark energy without matter}

First we will study the S-brane dark energy models  in the context of
pure dark energy without additional matter, just to get the feeling
about the relevant equation of state. Since we  know the scale factor
$S(\xi(t))$ we may find the equation of state of this kind of dark
energy 
\be
w= {p\over \rho}= - {1\over 3} -{2\over 3} {\ddot S S\over \dot
  S^2}=-{1- 2q\over 3} \ ,
\ee
where $q=-{\ddot S S\over \dot S^2}$ is the deceleration parameter and
$\dot S\equiv {dS\over d t}= S^{-3 }{dS\over d \xi}$. It can be also
given in terms of the $\xi$-time derivatives, $ S'\equiv {dS\over d
  \xi}$ 
\be
w= {5\over 3} -{2\over 3} { S^{''} S\over  (S')^2}
\ee
since an analytic expression for $S(\xi)$ is known. The function
$w(\xi)$ is plotted in Fig. 1. 

\begin{figure}[h!]
\centering\leavevmode\epsfysize=4.5cm \epsfbox{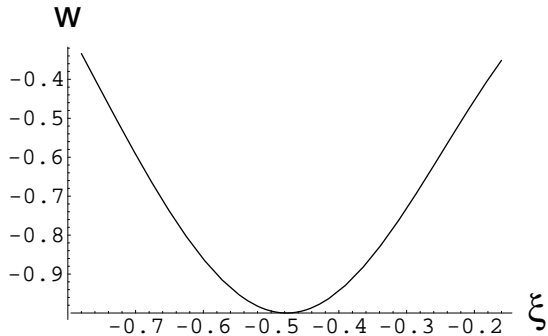}

\

\caption[fig1]
{Dark Energy equation of state $w(\xi)={5\over 3} -{2\over 3} { S^{''}
    S\over  (S')^2}$ 
corresponding to the S-brane solution $S(\xi)$ as the function of time $\xi$. }
\label{dark2}
\end{figure}
The plot of $w$ drops from the value $-1/3$ till $-1$ and afterwards
raises again towards  
$-1/3$. This is in agreement with the observation in
\cite{TW}-\cite{Emparan:2003gg} that the acceleration is changing 
the sign from negative to positive and back. Note also that one can
represent $w$ as a function of the kinetic energy of the scalar field
and a potential energy 
\be
w= {E_{kin}-V\over E_{kin}+V} \ .
\ee

In \cite{Emparan:2003gg} this property of the solution was interpreted
as follows: the field starts with large kinetic energy and the field
runs up the exponential hill, $w$ decreases. By the time all kinetic
energy is lost $w=-1$ is achieved when the field stops at some point
of the hill. It starts rolling down and $w$ increases. Before and
after the turning point the universe accelerates for some time. 

\subsection{The model including dark matter} 

To make this model realistic, one should add to
the energy momentum  tensor of the scalar field (dark energy) the
contribution of non-relativistic matter which  dominated the evolution
of the universe at earlier times.  
In the phenomenological setting, one can introduce matter by adding
the energy density of matter $\rho_M  \propto a^{-3}(t)$  to the
energy density of the scalar field. We will assume that the matter
energy density dominated in the early universe, leading to the
expansion of the universe with the Hubble constant $H={2\over 3 t}$. 
As a result, in the early universe the friction term
$3H\dot\phi$ in the equation for the field $\phi$ (see below) was very
large.   Therefore the field $\phi$ was sitting at the slope of the
potential and waiting until the Hubble parameter became sufficiently
small  
\cite{Kallosh:2002gf}. When the  
Hubble parameter $H\sim {2\over 3 t}$ becomes small enough, the field $\phi$ 
starts rolling down with vanishing initial velocity. This means that
the field $\phi$ evolves starting with $w=-1$ with increasing $w$.
In the language of  \cite{Emparan:2003gg}, the field $\phi$ evolves
\ from the position on the hill downwards only. 

We assume that the universe is spatially flat. Our equations are
\bea
\ddot\phi &+& 3H \dot \phi= -V_{,\phi}\\
H^2&=& {1\over 3} \left(\rho_M + {1\over 2}\dot \phi^2 + V(\phi)\right),
\label{Fried}\eea 
where $H={\dot a\over a}$ and 
\be
\rho_{\rm total}= \rho_M + {1\over 2}\dot \phi^2 + V(\phi) \ .
\ee
Density of matter satisfies the equation $\dot \rho_M + 3 H \rho_M=0$ which is 
solved by $\rho_M= {C\over a^3}$ where $C$ is a constant.
Here $a(t)$ is a scale factor on the universe with the metric
$ds^2= -dt^2 + a(t)^2d\vec x^2$. Note that $a(t)$ is not the same
function as $S(t)$ since eq. (\ref{Fried}) is different from
equations following from the 4d action (\ref{effective}) and therefore
different from the equations of 11d supergravity. 

\begin{figure}[h!]
\centering\leavevmode\epsfysize=4.5cm \epsfbox{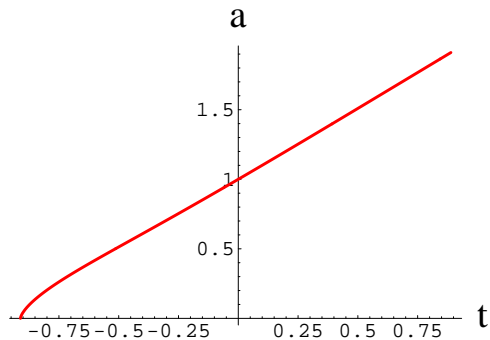}

\

\caption[fig1]
{The scale factor of the universe $a(t)$ as a function of the 4d
  observer time $t$.  The universe evolves from the Big Bang at
  $t\approx -1$  and $a$ close to zero. The point $t=0$,  defines
  ``today''where the scale factor is normalized to $1$. The future
  evolution of the scale factor is at $t>0$. In the model with
  potential (\ref{psipot}) it expands forever, approaching a Minkowski
  vacuum.  } 
\label{a}
\end{figure}

\begin{figure}[h!]
\centering\leavevmode\epsfysize=4.5cm \epsfbox{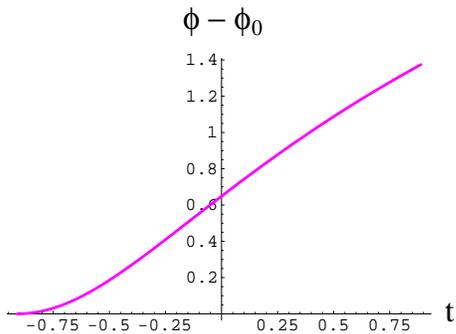}

\

\caption[fig1]
{The deviation of the scalar field from its initial value  $\phi_0$,
  in units $M_{Pl} =1$. The point $t=0$ corresponds to the present
  time.}  
\label{a2}
\end{figure}

We solved equations (\ref{Fried}) numerically. We found an initial
value $\phi_0$ of the field $\phi$, such that at the time when the
Hubble constant $H = {\dot a/a}$ reaches its presently observed value,
the fraction of the energy density concentrated in the field $\phi$
becomes equal to $\Omega_D = 0.7$. The method used to find such
solutions is explained in \cite{Kallosh:2002gf}. The corresponding
solutions for 
the scale factor of the universe $a(t)$, the deviation of the scalar
field from its initial value $\phi_0$, dark energy $\Omega_D(z)$,  and
equation of state $w(z)$ are plotted in Figs. 2, 3, 4, and 5.  Here
the redshift is defined via the scale factor as 
follows: $z= a^{-1}(t) -1$, and $a(t)$ is taken to be 1 at the present
time, corresponding to $z = 0$.

\begin{figure}[h!]
\centering\leavevmode\epsfysize=4.5cm \epsfbox{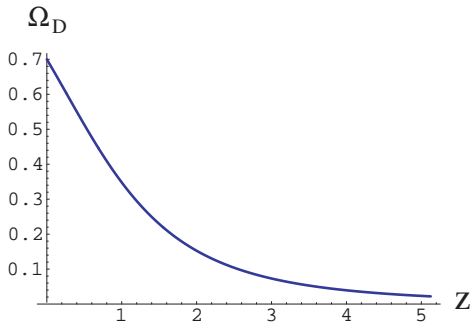}

\

\caption[fig1]
{The value of $\Omega_D= {\rho_D\over \rho_{\rm total}}$ raises from a
  very small value  at $z=5$ towards $0.7$ ``today'' at $z=0$. } 
\label{omega} 
\end{figure}

\begin{figure}[h!]
\centering\leavevmode\epsfysize=4.5cm \epsfbox{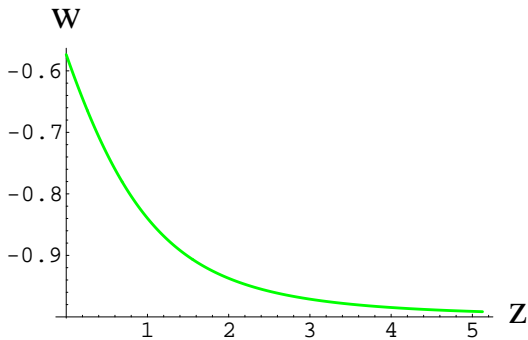}

\

\caption[fig1]
{The value of $w_D = {p_D\over \rho_{D}}$ raises from $\approx -1$  at
  $z\approx 5$ towards $-0.57$  at $z=0$.} 
\label{wz} 
\end{figure}

Note that one may also take for the value of $\Omega_D$ today $0.73$
which is one of the possibilities discussed in
\cite{Bennett:2003bz}. For this value of $\Omega_D$ we have also
evaluated the equations of state of the dark energy. It raises today
even more that the one for the $0.7$ case. 

The  observational constraint \cite{Bennett:2003bz} on a {\it
  time-independent} equation of state  is  $w< -0.78$. In our case $w$
remains smaller than $-0.78$ for $z >0.7$, but it grows up to $-0.57$
at $z=0$, which makes it rather vulnerable. The future cosmological
observations may either rule our this model or show that it is
compatible with the data.  

One should keep  in mind  that only the dark energy component of our
model was derived directly from M-theory in $d=11$. We have added `by
hand' the phenomenological  CDM component required for the consistency
of any cosmological model of late-time 
acceleration.  If not only dark energy but also  dark matter were derived
\ from M-theory, it could possibly affect some of our conclusions.

\section{Why $\Omega_D$ and $\Omega_M$ are of the same order of
  magnitude? A possible solution of the coincidence problem.} 

Finally, one should discuss the issue of fine-tuning of the initial
conditions in this model. We have found  cosmological solutions
consistent with the present observational data implying that 14
billion years after the big bang one has $\rho_D \sim 10^{-120}$ and
$\Omega_D \sim 0.7$. In order to find these solutions it was necessary
to  choose a proper initial value of the field $\phi = \phi_0$. One
can roughly estimate $\phi_0$ assuming that the field does not change
much during the cosmological evolution. This  leads to relation $21\,
e^{-\sqrt{18\over 7}\phi_0}\approx 10^{-120}$, and, consequently  
$\phi_0 \approx 174$, in Planck units. Numerical analysis shows that
the field $\phi$ grows only by $0.6$ during the cosmological evolution
up to the present time, see Fig. 3, so the expression $\phi_0 \approx
174$ gives a good estimate of the numerical magnitude of $\phi_0$.  

On the other hand, numerical analysis shows that if the initial value
of the field was equal to $\phi_0 -0.5$, then 14 billion years after
the big bang one would have $\Omega_D \sim 0.8$. Meanwhile, increasing
$\phi_0$ by $0.5$ would lead to $\Omega_D = 0.54$. Both values are
incompatible with cosmological observations. 

 Fixing initial value of the field $\phi$ with accuracy $\pm 0.5
 M_{Pl}$ may not seem to be such a terrible thing to do. However, if,
 in the absence of a better idea, we assume  that all initial values
 of $\phi$ are equally probable, we immediately realize that we have a
 serious fine-tuning problem. 

Indeed, if one takes initial value of $\phi$ in any place of the
semi-infinite interval  $\phi \gg \phi_0+ M_{Pl}$, one finds $\Omega_D
\ll 1$. Therefore one may argue that the probability to have $\Omega_D
\sim 0.7$ in this model is infinitesimally small. 

Another way to formulate this problem is to note that for any given
initial value of the field $\phi = \phi_0$, the cosmological time when
$\Omega_D$ becomes equal to $0.7$ is proportional to $H^{-1}(\phi_0)
\sim V^{-1/2}(\phi_0) \sim e^{\sqrt{9\over 14}\phi_0}$. Thus the
cosmological time when $\Omega_D$ becomes equal to $0.7$, and, more
generally, the time when $\Omega_D$ and $\Omega_M$ are of the same
order of magnitude, is exponentially sensitive to initial
conditions. Therefore it looks very surprising that we live at the
time when $\Omega_D$ and $\Omega_M$ are of the same order of magnitude
(the coincidence problem).

This problem is quite generic for most of the models of dark
energy. However, there are some models that do not suffer from this
problem. The simplest (and perhaps the first) model of dark energy was
proposed in \cite{Linde86}. It has a linear potential $V(\phi) =\alpha
\phi + C$. If the slope of the potential is sufficiently small (a
condition that should be satisfied in all versions of models of dark
energy), the field $\phi$ practically does not move during the last 14
billions of years, so the potential acts as an effective cosmological
constant. During eternal inflation, the universe becomes divided into
many different exponentially large domains with different values of
$\phi$, and, correspondingly, with different values of the effective
cosmological constant \cite{Linde86}.  In those parts of the universe
where $V(\phi) < -10^{-119}$, the universe collapses within the time
smaller than 14 billion years \cite{Barrow,Linde86,Kallosh:2002gg}. In
those parts of the universe where $V(\phi) \gg 10^{-119}$, the
probability of formation of galaxies would be strongly suppressed
\cite{Linde86,Weinberg87,GarrigaVil}. Thus, in the context of this
simple model, we can live and make our observations only in those
parts of the universe where $|V(\phi)| \lesssim 10^{-119}$, which
corresponds to a finite range of values of the field $\phi$
\cite{Linde86}.  

A  different  anthropic solution of the cosmological constant problem
and the coincidence problem  was proposed  for the dark energy model
based on N=8 supergravity \cite{Kallosh:2002gf,Kallosh:2002gg}. In
that model, the universe collapses very fast, unless the present value
of the effective cosmological constant is smaller than $10^{-119}$. 

One could try to use similar arguments  for the dark energy model
discussed above. Indeed, we would be unable to live in a universe with
initial value of the field $\phi < \phi_0 -3$  because for small
$\phi$ one has $V(\phi) \gg 10^{-119}$. Therefore galaxies would not
form in such a universe, in accordance with
\cite{Linde86,Weinberg87,GarrigaVil}. In fact, the situation here is
even better than in the models studied
\cite{Linde86,Weinberg87,GarrigaVil}. Indeed, in these models the
probability distribution was supposed to be constant with respect to
the vacuum energy. Therefore the suppression  of galaxy formation for
$\rho_D \gg 10^{-119}$ did not provide an entirely satisfactory
explanation of the present value of the vacuum energy $\rho_D \sim
10^{-120}$. Meanwhile, in our case the situation is somewhat better
because the canonically normalized variable is not $V(\phi)$ but
$\phi$. The energy density $V(\phi)$ depends on $\phi$ exponentially,
so the change of $V(\phi)$ by one or two orders of magnitude occurs
within a small range of variation of the field $\phi$. If the
probability is constant with respect to the field   $\phi$ (which is a
reasonable assumption in the context of inflationary cosmology
\cite{Linde86,GarrigaVil}), then the  probability to live in a part of
the universe with $V(\phi) \gg  10^{-120}$ becomes much stronger
suppressed than in the models studied in
\cite{Linde86,Weinberg87,GarrigaVil}. 

However, at the first glance, in this model we must live in a universe
with $V(\phi) \ll  10^{-120}$.  Indeed, galaxies can be formed and the
universe does not collapse  for all of the initial values  $\phi
\gtrsim \phi_0$. The probability to live in  a universe with the field
$\phi$ belonging to the finite interval  $-0.5<  \phi-\phi_0 <0.5$
seems to be infinitesimally small as compared with the probability to
live in a universe in the  infinitely large interval $\phi \gg
\phi_0$, with dark energy $\Omega_D \ll 1$. 

One can solve this problem if one finds a true 11d implementation of
this model, as discussed in Section II. Indeed, let us remember that
the mass gap for the KK modes in our model is given by  
\be
m_{\rm KK}  \sim c\ \sqrt V(\phi)  \sim c\ e^{-\sqrt{9\over 14}\phi}
\sim c\ e^{-0.8\phi} \ . 
\ee
If $c = O(1)$, then, as we argued, the theory is effectively 11d, and
the model does not work \cite{E2}. Let us assume for a moment that
some resolution of this problem will be found, so that  the KK mass
gap is large enough for $\phi \sim \phi_0$. However, this gap
exponentially rapidly disappears for $\phi \gg \phi_0 $  and therefore
our world becomes effectively 11d for $\phi \gg \phi_0 $. For example,
even if the mass gap at $\phi \sim \phi_0$ is as large as $M_{Pl} =1$,
it becomes much smaller than 1 TeV and our universe becomes
effectively 11d for $\phi -\phi_0\gtrsim 45$. If the mass gap at $\phi
\sim \phi_0$  is about 1 TeV, the universe becomes effectively 11d for
$\phi -\phi_0 \gg 1$. 

Life as we know it can exist  only under  fine-tuned relations between
masses and coupling constants, which would be strongly affected by
appearance of new low-mass states \cite{Barrow}. This suggests that
life of our type may exist only if the deviation $\Delta \phi =\phi
-\phi_0$ of the initial value of the field $\phi$ from $\phi_0 \sim
174$ was very limited, $-3< \Delta \phi < O(10)$. Therefore it is not
very surprising that we live in the universe with $-0.5< \Delta \phi
<0.5$. This provides a possible resolution of the cosmological
coincidence problem discussed above. 

The structure of the universe in this model looks especially
interesting if one can incorporate it into the eternal inflation
scenario \cite{Vil83,Lin86}. If the potential of the field $\phi$
remains very flat during inflation, then inflationary quantum
fluctuations of the  field $\phi$ divide the universe into infinitely
large number of exponentially large domains with all possible values
of $\phi$ being approximately equally represented, just like in the
dark energy models  of Refs. \cite{Linde86,GarrigaVil}. The domains
with $\phi \ll \phi_0$ will be effectively 4d, but they will contain
no galaxies. The domains with $\phi \gg \phi_0$ will be effectively
11d, unsuitable for life as we know it. We can live only in domains
with $\phi \sim \phi_0$. In a considerable part of such domains
$\Omega_D$ and $\Omega_M$ are of the same order of magnitude. In a
distant future, the value of the field $\phi$ in each of these domains
will grow, the mass of the KK model will fall down exponentially, the
4d space will gradually decompactify and become 11d. 

\

As we already emphasized, the model of dark energy discussed in our
paper suffers from many serious problems.  In this respect, this model
seems much less satisfactory than the theory of cosmic acceleration in
a metastabe de Sitter state in a context of string theory with
stabilized moduli \cite{Kachru:2003aw}.  It may still be interesting
that the model discussed above provides a possibility to describe the
universe marginally consistent with the present observational
data. Some of the features of this model may have a more general
significance. In particular, the link between the vacuum energy
density and the effective dimensionality  may appear in other models
as well. It may provide a new way towards a solution of the
cosmological constant problem.

We are grateful to R.~Emparan, J.~Garriga, L. Susskind, and
E. Silverstein for useful comments. 
This work was supported by NSF grant PHY-9870115. The work by A.L. was
also supported 
by the Templeton Foundation grant No. 938-COS273.


\begin{thebibliography}{4}

\bibitem{Bennett:2003bz}
C.~L.~Bennett {\it et al.},
``First Year Wilkinson Microwave Anisotropy Probe (WMAP) Observations:
Preliminary Maps and Basic Results,'' 
arXiv:astro-ph/0302207.



\bibitem{Kachru:2003aw}
S.~Kachru, R.~Kallosh, A.~Linde and S.~P.~Trivedi,
``De Sitter vacua in string theory,''
arXiv:hep-th/0301240.

\bibitem{Gutperle:2002ai}
M.~Gutperle and A.~Strominger,
``Spacelike branes,''
JHEP {\bf 0204} (2002) 018
[arXiv:hep-th/0202210].


\bibitem{cgg}
C.~M.~Chen, D.~V.~Gal'tsov and M.~Gutperle,
``S-brane solutions in supergravity theories,''
Phys.\ Rev.\ D {\bf 66} (2002) 024043 [arXiv:hep-th/0204071].

\bibitem{kmp}
M.~Kruczenski, R.~C.~Myers and A.~W.~Peet,
``Supergravity S-branes,''
JHEP {\bf 0205} (2002) 039 [arXiv:hep-th/0204144];

\bibitem{ohta1}
N.~Ohta,
``Intersection rules for S-branes,''
Phys.\ Lett.\ B {\bf 558} (2003) 213 [arXiv:hep-th/0301095].


\bibitem{sbunch}
S.~Roy,
``On supergravity solutions of space-like Dp-branes,''
JHEP {\bf 0208} (2002) 025 [arXiv:hep-th/0205198];
N.~S.~Deger and A.~Kaya,
``Intersecting S-brane solutions of D = 11 supergravity,''
JHEP {\bf 0207} (2002) 038 [arXiv:hep-th/0206057].



\bibitem{TW}
P.~K.~Townsend and M.~N.~Wohlfarth,
``Accelerating cosmologies from compactification,''
arXiv:hep-th/0303097.


\bibitem{ohta}
N.~Ohta,
``Accelerating cosmologies from S-branes,''
arXiv:hep-th/0303238.

\bibitem{roy}
S.~Roy,
``Accelerating cosmologies from M/string theory compactifications,''
arXiv:hep-th/0304084.


\bibitem{wolf}
M.~N.~Wohlfarth,
``Accelerating cosmologies and a phase transition in M-theory,''
arXiv:hep-th/0304089.

\bibitem{Emparan:2003gg}
R.~Emparan and J.~Garriga,
``A note on accelerating cosmologies from compactifications and S-branes,''
arXiv:hep-th/0304124.

\bibitem{Ohta:2003ie}
N.~Ohta,
``A Study of Accelerating Cosmologies from Superstring/M theories,''
arXiv:hep-th/0304172.

\bibitem{Chen:2003ij}
C.~M.~Chen, P.~M.~Ho, I.~P.~Neupane and J.~E.~Wang,
``A Note on Acceleration from Product Space Compactification,''
arXiv:hep-th/0304177.






\bibitem{hyperb}
N.~Kaloper, J.~March-Russell, G.~D.~Starkman and M.~Trodden,
``Compact hyperbolic extra dimensions: Branes, Kaluza-Klein modes and
cosmology,'' 
Phys.\ Rev.\ Lett.\  {\bf 85} (2000) 928 [arXiv:hep-ph/0002001].



\bibitem{Atiyah:2001qf}
M.~Atiyah and E.~Witten,
``M-theory dynamics on a manifold of G(2) holonomy,''
Adv.\ Theor.\ Math.\ Phys.\  {\bf 6}, 1 (2003)
[arXiv:hep-th/0107177].

\bibitem{Witten:2001uq}
E.~Witten,
``Anomaly cancellation on G(2) manifolds,''
arXiv:hep-th/0108165.

\bibitem{Acharya:2001gy}
B.~Acharya and E.~Witten,
``Chiral fermions from manifolds of G(2) holonomy,''
arXiv:hep-th/0109152.



\bibitem{Kallosh:2002gf}
R.~Kallosh, A.~Linde, S.~Prokushkin and M.~Shmakova,
``Supergravity, dark energy and the fate of the universe,''
Phys.\ Rev.\ D {\bf 66}, 123503 (2002)
[arXiv:hep-th/0208156].



\bibitem{E1} {Expression for $V(M)$ obtained  in \cite{Emparan:2003gg}
    was presented there for $r_c = 1$. However, for the subsequent
    analysis we must keep $r_c$ as a free parameter. Restoring $r_c$
    in this expression, Eq. (\ref{psipot}), leads to  modification of
    some of the results obtained in the previous version of our
    paper. We are grateful to Emparan and Garriga for pointing this
    out.} 

\bibitem{LopesFranca:2002ek}
U.~J.~Lopes Franca and R.~Rosenfeld,
``Fine tuning in quintessence models with exponential potentials,''
JHEP {\bf 0210}, 015 (2002)
[arXiv:astro-ph/0206194].

\bibitem{E2} { We are thankful to Emparan and Garriga for a detailed
    discussion of this issue.} 

\bibitem{Appelquist:1982zs}
T.~Appelquist and A.~Chodos,
``Quantum Effects In Kaluza-Klein Theories,''
Phys.\ Rev.\ Lett.\  {\bf 50} (1983) 141.

\bibitem{Appelquist:1983vs}
T.~Appelquist and A.~Chodos,
``The Quantum Dynamics Of Kaluza-Klein Theories,''
Phys.\ Rev.\ D {\bf 28} (1983) 772.


\bibitem{Kehagias:2000dg}
A.~Kehagias and J.~G.~Russo,
``Hyperbolic spaces in string and M-theory,''
JHEP {\bf 0007} (2000) 027
[arXiv:hep-th/0003281].

\bibitem{Nasri:2002rx}
S.~Nasri, P.~J.~Silva, G.~D.~Starkman and M.~Trodden,
``Radion stabilization in compact hyperbolic extra dimensions,''
Phys.\ Rev.\ D {\bf 66} (2002) 045029
[arXiv:hep-th/0201063].

\bibitem{Hull:jw}
C.~M.~Hull and N.~P.~Warner,
``Noncompact Gaugings From Higher Dimensions,''
Class.\ Quant.\ Grav.\  {\bf 5}, 1517 (1988).



\bibitem{Linde86}  A.D. Linde, ``Inflation And Quantum Cosmology,''
Print-86-0888
(June 1986),
in: {\it  Three hundred years of gravitation}, (Eds.: Hawking, S.W.
and Israel, W., Cambridge Univ. Press, 1987), 604-630. 

\bibitem{Barrow}   J. D. Barrow and F. J. Tipler  {\it The Anthropic
    Cosmological Principle}, (Oxford University Press, New York,
  1986). 

\bibitem{Kallosh:2002gg}
R.~Kallosh and A.~Linde,
``M-theory, cosmological constant and anthropic principle,''
Phys.\ Rev.\ D {\bf 67}, 023510 (2003)
[arXiv:hep-th/0208157].

\bibitem{Weinberg87} S. Weinberg, Phys. Rev. Lett. {\bf 59}, 2607 (1987);
 G. Efstathiou, M.N.R.A.S. {\bf 274},
L73 (1995);
A.~Vilenkin,
``Quantum Cosmology and the Constants of Nature,''
arXiv:gr-qc/9512031; 
H.~Martel, P.~R.~Shapiro and S.~Weinberg,
``Likely Values of the Cosmological Constant,'' Astrophys. J. {\bf
  492}, 29 (1998) 
arXiv:astro-ph/9701099;
S.~A.~Bludman and M.~Roos,
``Quintessence cosmology and the cosmic coincidence,''
Phys.\ Rev.\ D {\bf 65}, 043503 (2002)
[arXiv:astro-ph/0109551].

\bibitem{GarrigaVil} J.~Garriga and A.~Vilenkin,
``Solutions to the cosmological constant problems,''
Phys.\ Rev.\ D {\bf 64}, 023517 (2001)
[arXiv:hep-th/0011262];
J.~Garriga and A.~Vilenkin,
``Testable anthropic predictions for dark energy,''
Phys.\ Rev.\ D {\bf 67}, 043503 (2003)
[arXiv:astro-ph/0210358]; A.~Linde,
``Inflation, quantum cosmology and the anthropic principle,''
arXiv:hep-th/0211048.


\bibitem{Vil83} A. Vilenkin,  
``The Birth Of Inflationary Universes,''
Phys.\ Rev.\ D {\bf 27}, 2848 (1983).

\bibitem{Lin86} A. D. Linde,  
``Eternally Existing Self-reproducing Chaotic Inflationary Universe,''
Phys.\ Lett.\ B {\bf 175}, 395 (1986);
A. D. Linde  and M.~I. Zelnikov, 
``Inflationary Universe With Fluctuating Dimension,''
Phys.\ Lett.\ B {\bf 215}, 59  (1988); A. D. Linde,   {\it Particle
  Physics and Inflationary Cosmology} (Harwood  
Academic Publishers, Chur, Switzerland, 1990); A. D. Linde,
D.A. Linde,    and A. Mezhlumian,   
``From the Big Bang theory to the theory of a stationary universe,''
Phys.\ Rev.\ D {\bf 49}, 1783 (1994)
[arXiv:gr-qc/9306035].



\end{thebibliography}
\end{document}